\documentclass[11pt,a4paper]{article}
\pdfoutput=1

\usepackage{cite}
\usepackage{physics}
\usepackage[english]{babel}
\usepackage[utf8]{inputenc}
\usepackage{amsmath}
\usepackage{graphicx}
\usepackage[nottoc]{tocbibind}
\usepackage{amssymb}
\usepackage{mathtools}

\usepackage[colorlinks=true, allcolors=blue]{hyperref}
\usepackage{float}
\usepackage{multirow}
\usepackage{makeidx}

\numberwithin{figure}{section}
\numberwithin{equation}{section}

\newcommand{\be}{\begin{equation}}
\newcommand{\ee}{\end{equation}}
\newcommand{\bea}{\begin{eqnarray}}
\newcommand{\eea}{\end{eqnarray}}
\def\beal#1\eeal{\begin{align}#1\end{align}}   
\def\besp#1\eesp{\begin{multline}#1\end{multline}} 

\newcommand\ie{\textit{i.e.}\ }
\newcommand\eg{\textit{e.g.}\ }
\newcommand\cf{\textit{cf.}\ }

\newcommand{\aka}{{a.k.a.}\ }

\newcommand{\viz}{{\it viz.}\ }

\newcommand{\ph}{\varphi}
\newcommand{\half}{\tfrac{1}{2}}
\newcommand{\e}{\text{e}}

\newcommand{\cO}{\mathcal{O}}

\begin{document}
\begin{titlepage}

\begin{center}
{\huge \bf Universal scaling dimensions for highly irrelevant operators in the Local Potential Approximation}


\end{center}
\vskip1cm


\begin{center}
{\bf Vlad-Mihai Mandric, Tim R. Morris and Dalius Stulga}
\end{center}

\begin{center}
{\it STAG Research Centre \& Department of Physics and Astronomy,\\  University of Southampton,
Highfield, Southampton, SO17 1BJ, U.K.}\\
\vspace*{0.3cm}
{\tt  V.M.Mandric@soton.ac.uk, T.R.Morris@soton.ac.uk, D.Stulga@soton.ac.uk}
\end{center}

\begin{abstract}
We study $d$-dimensional scalar field theory in the Local Potential Approximation of the functional renormalization group. Sturm-Liouville methods allow the eigenoperator equation to be cast as a Schr\"odinger-type equation. Combining solutions in the large field limit with the Wentzel–Kramers–Brillouin approximation, we solve analytically for the scaling dimension $d_n$ of high dimension potential-type operators $\cO_n(\ph)$ around a non-trivial fixed point. We find that $d_n = n(d-d_\ph)$ to leading order in $n$ as $n\to\infty$, where $d_\ph=\frac12(d-2+\eta)$ is the scaling dimension of the field, $\ph$, and determine the power-law growth of the subleading correction.
For $O(N)$ invariant scalar field theory, the scaling dimension is just double this, for all fixed $N\ge0$ and additionally for $N=-2,-4,\ldots \,.$ 
These results are universal, independent of the choice of cutoff function which we keep general throughout, subject only to some weak constraints.
\end{abstract}


\end{titlepage}

\newpage

\tableofcontents

\newpage

\section{Introduction}

The functional renormalization group (FRG) is one of the most widely used approaches to study quantum field theories in non-perturbative regimes, as evidenced by an extensive literature (see, for instance, the reviews \cite{Morris:1998,Bagnuls:2000,Berges:2000ew,Pelissetto:2000ek,Rosten:2010vm,Dupuis:2020fhh}).
Various realizations of the FRG exist \cite{Wilson:1973,Wegner:1972ih,Polchinski:1983gv,Latorre:2000qc,Arnone:2002cs,Nicoll1977,Wetterich:1992,Morris:1993,Weinberg:1976xy,Bonini:1992vh,Ellwanger1994a,Morgan1991}, but the most prevalent version \cite{Nicoll1977,Wetterich:1992,Morris:1993,Weinberg:1976xy,Bonini:1992vh,Ellwanger1994a,Morgan1991} focuses on the flow of an appropriately defined Legendre effective action $\Gamma_\Lambda$ (also referred to as the effective average action), with respect to an infrared cut-off scale $\Lambda$. This flow equation is given by:
\begin{equation}
\label{flow}
    \frac{\partial}{\partial\Lambda} \Gamma_\Lambda=-\frac{1}{2}\text{Tr}\left[   \frac{1}{\Delta_\Lambda} \frac{\partial \Delta_\Lambda}{\partial\Lambda}\left(1+\Delta_\Lambda \Gamma^{(2)}_\Lambda\right)^{-1}  \right] \,.
\end{equation}
Here, Tr stands for a space-time trace and $\Gamma^{(2)}_\Lambda$ is the Hessian with respect to the fields. The propagator $\Delta_\Lambda(q)=C_\Lambda(q)/q^2$ is modified by the inclusion of a multiplicative infrared cutoff function $C_\Lambda(q)=C(q^2/\Lambda^2)$, which is non-negative, monotonically increasing, and satisfies $C(0)=0$ and $C(\infty)=1$.

In practical applications, some form of approximation becomes necessary. One frequently employed approximation is the Local Potential Approximation (LPA) \cite{Nicoll:1974zz,Hasenfratz:1985dm,Felder:1987,Ball:1994ji,Morris:1994ki,Morris:1995af,Aoki:1996fn,Comellas:1997tf,Osborn:2009vs,Bervillier:2013hha}, which simplifies the flow equations by disregarding the momentum dependence of the effective action, except for a local potential term, $V_\Lambda$. For a scalar field $\ph$ in $d$ Euclidean dimensions, the effective action then takes the form:
\begin{equation}
    \Gamma_\Lambda=\int\!\! d^dx \left(\frac{1}{2}(\partial_\mu \ph)^2+ V_\Lambda(\ph)\right) \,.
\end{equation}
While an exact analytical solution to this truncated FRG formulation is still not possible in general, the LPA enables numerical treatments that provide valuable insights into the system's behaviour. It allows for numerical estimates of various physical quantities, including critical exponents and the scaling equation of state \cite{Morris:1994ki,Morris:1994ie,Morris:1996xq,Morris:1997xj,Morris:1998,Bagnuls:2000,Berges:2000ew,Pelissetto:2000ek,Rosten:2010vm,Dupuis:2020fhh}. Moreover, the LPA serves as the initial step in a systematic derivative expansion \cite{Morris:1994ki,Morris:1994ie,Morris:1996xq,Morris:1997xj,Morris:1998}, which facilitates a more comprehensive exploration of the system's properties \cite{Morris:1996xq,Morris:1998,Bagnuls:2000,Berges:2000ew,Pelissetto:2000ek,Rosten:2010vm,Dupuis:2020fhh}.

Nevertheless it is important to acknowledge the limitations of the LPA and more generally the derivative expansion. Since such truncations do not correspond to a controlled expansion in some small parameter, the errors incurred can be expected to be of the same order in general as the quantities being computed\footnote{See however refs. \cite{DePolsi:2020pjk,DePolsi:2021cmi,DePolsi:2022wyb,Balog:2019rrg}}. Furthermore, quantities that should be universal, and thus independent of the specific form of the cutoff, are not (for example for the critical exponent $\nu$ at the Wilson-Fisher fixed point in $d=3$ dimensions,  the LPA yields $\nu=0.689$ with a sharp cutoff \cite{Morris:1994ki} whilst for a power-law cutoff one obtains $\nu=0.660$ \cite{Morris:1994ie}).

It has long been understood that an exception to this is the general form of a non-trivial fixed potential $V(\ph)$ in the large field regime \cite{Morris:1994ki,Morris:1994ie,Morris:1997xj,Morris:1998}, which follows from asymptotic analysis:
\be \label{asympV} V(\ph) = A|\ph|^{d/d_\ph}+\cdots\qquad\text{as}\qquad \ph\to\pm\infty\,, \ee
where the ellipses stand for subleading terms (see later).
The leading term coincides with the scaling equation of state precisely at the fixed point. It is a simple consequence of dimensional analysis on using the scaling dimension $d_\ph =\frac12(d-2+\eta)$ for the field $\ph$ at the fixed point, $\eta$ being its anomalous dimension. However asymptotic analysis does not fix the amplitude
$A$ or the anomalous dimension $\eta$, which have to be found by other means, for example by numerical solution of truncated fixed point equations. 

In this paper, we will show that within LPA, asymptotic analysis combined with Sturm-Liouville (SL) and Wentzel–Kramers–Brillouin (WKB) analysis,\footnote{See \eg ref. \cite{Ince1956} for textbook discussion of SL methods and ref. \cite{morse1953methods} for WKB methods.} also allows one to determine asymptotically the scaling dimension $d_n$ of the highly irrelevant ($d_n\gg1$) eigenoperators $\cO_n=\cO_n(\ph)$ of potential-type (those containing no spacetime derivatives). Ordering them by increasing scaling dimension, we will show that $d_n = n(d-d_\ph)$ to leading order in $n$. In the case of $O(N)$ invariant scalar field theory with fixed $N\ge0$ the dimension $d_n$ is doubled to $d_n = 2n(d-d_\ph)$. The scaling dimension is thus independent of $N$. It agrees with the result for the single scalar field since these eigenoperators are functions of $\ph^2=\ph^a\ph^a$, and thus pick out only the even eigenoperators (those symmetric under $\ph\leftrightarrow-\ph$) in the $N=1$ case. We also show that the scaling dimension is $d_n = 2n(d-d_\ph)$ whenever $N=-2k$, where $k$ is a non-negative integer. 


Once again these results are independent of the choice of cutoff and thus universal. Indeed in this paper, we will keep the cutoff function completely general throughout, subject only to some weak technical constraints that we derive later. Note that, like the fixed point equation of state \eqref{asympV}, the $d_n$ take the same form, independent of the choice of fixed point, provided only that $d_\ph>0$ and that the fixed point potential is non-vanishing. 
We also show that the next to leading correction to $d_n$ behaves as a power of $n$. The power is universal although the coefficient of the subleading correction is not.


Actually this approach was first employed to determine the scaling dimension of highly irrelevant eigenoperators in an $f(R)$ approximation \cite{Mitchell:2021qjr,Morris:2022btf} to the asymptotic safety scenario \cite{Reuter:1996,Percacci:2017fkn,Reuter:2019byg} in quantum gravity. 
The $f(R)$ approximation serves as a close analogue to the LPA in this context \cite{Benedetti:2012,Dietz:2012ic,Benedetti:2013jk}.  However, while the resulting scaling dimensions $d_n$ exhibit a simple nearly-universal form for large values of $n$, they nevertheless retained strong dependence on the choice of cutoff. This issue can be traced back \cite{Morris:2022btf} to the so-called single-metric (or background field) approximation \cite{Reuter:1996}, where  the identification of the quantum metric with the background metric is made in order to close the equations. The present paper thus completes the circle by demonstrating that, indeed, without such an approximation, the results become truly universal. Additionally, it showcases the power of these methods in a simpler context.

The paper is organised as follows. We first analyse the functional renormalization group equations for a single scalar field in the LPA. From the eigenoperator equation we write the resulting SL equation in Schr\"odinger form and thus, by taking the large field limit, deduce the asymptotic form of the renormalization group eigenvalues in the WKB limit. Sec. \ref{sec:on} extends the analysis to $O(N)$ scalar field theory using the same approach. Finally in sec. \ref{sec:summary} we conclude and discuss the results, placing them in a wider context.

\section{Flow equations in LPA}
The LPA approximation amounts to setting the field $\ph$ in the Hessian to a spacetime constant, thus dropping from a derivative expansion all terms that do not take the form of a correction to the potential. The flow equation for $V_\Lambda(\ph)$ then takes the form:
\begin{equation}
    \left( \partial_t + d_\varphi \varphi \frac{\partial}{\partial \varphi}-d \right) V_\Lambda(\varphi)=-\frac12\int \frac{d^d q}{(2\pi)^d}\frac{\dot{\Delta}}{\Delta}\frac{1}{1+\Delta V''_\Lambda(\varphi)}\,, \label{flowLPA}
\end{equation}
where $\partial_t=-\Lambda\partial_\Lambda$, $t$ being the renormalization group `time' which, following \cite{Wilson:1973}, we have chosen to flow towards the IR. Here the momentum, potential and field are already scaled by the appropriate power of $\Lambda$ to make them dimensionless. Then $\Delta = C(q^2)/q^2$ no longer depends on $\Lambda$. The same is true of $\partial_t \Delta_\Lambda$, which after scaling we write as $\dot{\Delta}$, where
\be 
\dot{\Delta} =  2\, C'(q^2)\,.
\ee
Since $C(q^2)$ is monotonically increasing, we have that $\dot{\Delta}>0$. 

The scaling dimension of the field is $d_\varphi=\frac{1}{2}(d-2+\eta)$, where $\eta$ is the anomalous dimension. Since $\eta$ arises from the renormalization group running of the field, and is typically inferred from corrections to the kinetic term, one would naturally conclude that it vanishes in LPA \cite{Wilson:1973,Nicoll:1974zz,Hasenfratz:1985dm,Felder:1987,Ball:1994ji,Morris:1994ki,Morris:1995af,Aoki:1996fn,Comellas:1997tf,Bagnuls:2000}. Nevertheless, as noticed in refs. \cite{Osborn:2009vs,Bervillier:2013hha}, this assumption is not necessary. The flow equation \eqref{flowLPA} is still a mathematically consistent equation with $\eta\ne0$. However, since we cannot determine $\eta$ directly from \eqref{flowLPA}, its value needs to be input from elsewhere (either from experiment or other theoretical studies). We will follow this strategy, in the expectation that it improves the accuracy of our final estimates for $d_n$.

Let us recall that the flow equation \eqref{flowLPA} is an implementation of the Wilsonian RG \cite{Wilson:1973,Morris:1998}. Lowering the cutoff $\Lambda$ implements the Kadanoff blocking \cite{Kadanoff:1966wm}, whilst rescaling the cutoff back to the original size is equivalently implemented by `measuring' all quantities in units of $\Lambda$ \ie by making them dimensionless using the appropriate power of $\Lambda$ \cite{Morris:1998} as we have done above. Then at a critical point corresponding to a continuous phase transition, the solutions $V_\Lambda(\ph)$ remain finite but the distinguishing feature is that they become independent of $\Lambda$ (see \eg \cite{Morris:1998}).

Thus at such a FP (fixed point) $V_\Lambda (\varphi)=V(\varphi)$, and $\eta$, have no renormalization group time dependence. The eigenoperator equation follows from linearising about a FP:
\begin{equation}
\label{eigenop}
    V_\Lambda(\varphi)= V(\ph)+\varepsilon\, v(\ph) \,\e^{\lambda t} \,,
\end{equation}
$\varepsilon$ being infinitesimal. Here $\lambda$ is the RG eigenvalue.  It is the scaling dimension of the corresponding coupling, and is positive (negative) for relevant (irrelevant) operators. The scaling dimension of the operator $v(\ph)$ itself is then $d-\lambda$. We write the eigenoperator equation in the same form as refs. \cite{Benedetti:2013jk,Mitchell:2021qjr,Morris:2022btf}:
\begin{equation}
    -a_2(\ph)v''(\ph)+a_1(\ph)v'(\ph)+a_0(\ph)v(\ph)=(d-\lambda)v(\ph)\,, \label{eig}
\end{equation}
where the $\ph$-dependent coefficients multiplying the eigenoperators are given by: 
\begin{align}
    a_0(\ph) &= 0 \,, \\
    a_1(\ph) &= d_\ph \ph \,,\label{a1} \\
    a_2(\ph) &= \frac12\int \frac{d^d q}{(2\pi)^d}\frac{\dot{\Delta}}{(1+\Delta V'')^2} >0\,, \label{a2}
\end{align}
and we have noted that $a_2$ is positive. 
We can now repeat the analysis carried out in \cite{Benedetti:2013jk,Mitchell:2021qjr,Morris:2022btf} to solve for $\lambda$ in the case of high dimension eigenoperators. 

\subsection{Asymptotic solutions}
For large $\varphi$, the RHS of \eqref{flowLPA} can be neglected. Thus at a fixed point, the equation reduces to a first order ODE (ordinary differential equation) which is easily solved. It gives the first term \eqref{asympV} in an asymptotic series solution \cite{Morris:1994ie}:
\begin{equation}
    V(\varphi)= A |\ph|^{m}+O\left(|\ph|^{2-m}\right) \quad \text{as} \quad \ph \rightarrow \pm \infty \,, \label{sol}
\end{equation}
where for convenience we introduce 
\be m = {d}/{d_\ph}\,, \ee 
and $A$ is a real constant (that is determined by solving for the full FP solution). The subleading terms arise from iterating the leading order contribution to next order.

Of course there is always the trivial $V(\ph)\equiv0$ fixed point solution, corresponding to the Gaussian fixed point. We will not be interested in that  (the scaling dimensions in that case are exactly known and reviewed in the discussion in sec. \ref{sec:summary}). Instead we focus on non-trivial FP solutions for which $A\ne0$. In principle, $A$ could be different in the two limits $\ph\to\pm\infty$, although in practice the fixed point potentials \eqref{sol} are symmetric. Anyway, we will see that $A$ drops out of the analysis in a few further steps.  

It is helpful for the following to note that $m>3$, since this inequality ensures that the $m$-dependent asymptotic solutions we are about to derive, are valid. To see that $m>3$, first note that if $\eta$ is neglected (typically $\eta\ll1$, see \eg \cite{ZinnJustin:2002ru}),  $m$ is a decreasing function of $d$ for all $d>2$. In practice, non-trivial FP solutions only exist for $2\le d<4$ (see \eg \cite{Morris:1994ki}). In the limit $d\to4^-$, $\eta\to0$ (by the $\epsilon$ expansion \cite{ZinnJustin:2002ru}) and thus $m\to4$. Therefore, if we can neglect $\eta$, we see that $m$ is bounded below by $m\ge 4$. In practice one finds that the values of $\eta$ increase as $d$ is lowered, but even in $d=2$ dimensions they are not large enough to destroy this bound. 
In $d=2$ dimensions, the asymptotic solution \eqref{sol} corresponds to that of a unitary minimal model \cite{Zamolodchikov:1986db,Morris:1994jc}. The one with the largest anomalous dimension is that of the Ising model universality class which has $\eta=1/4$, thus in $d=2$ dimensions we have in fact $m\ge8$ for all the unitary minimal models. In this way, we see that we are safe to bound $m>3$ in practice.

Note that the solution \eqref{sol} has a single free parameter even though the FP equation is a (non-linear) second order ODE. The second parameter, if it exists, can be deduced by linearising around \eqref{sol}, writing $V(\ph)\mapsto V(\ph)+\delta V(\ph)$, and solving the flow equation \eqref{flowLPA} at the FP this time for $\delta V$. Since $\delta V$ satisfies a \emph{linear} second order ODE and one solution is already known, namely $\delta V = \partial_A V(\ph)$, it is easy to find the solution that corresponds at the linearised level to the missing parameter \cite{Morris:1994ki,Morris:1994ie}.  However, one then discovers that these `missing' linearised solutions are rapidly growing exponentials. Such a linearised perturbation is not valid asymptotically since for diverging $\ph$ it is much larger than the solution \eqref{sol} we perturbed around. Hence, the FP asymptotic solutions only have the one free parameter, $A$. 

Substituting \eqref{sol} into \eqref{a2}, we see that asymptotically $a_2(\ph)$ scales as follows:
\begin{equation}
    a_2(\ph) = F\,|\ph|^{2(2-m)} + O\left( |\ph|^{3(2-m)}\right) \quad \text{as} \quad \ph \rightarrow \pm \infty \,, \label{limA2}
\end{equation}
where $F$ is positive and cutoff dependent:
\begin{equation}
\label{F}
   F = \frac{1}{2\left(m(m-1)A\right)^2}  \int \frac{d^d q}{(2\pi)^d}\frac{\Dot{\Delta}}{\Delta^2} 
   = - \frac{1}{\left(m(m-1)A\right)^2} \int \frac{d^d q}{(2\pi)^d} \,q^4\frac{\partial}{\partial q^2} C^{-1}(q^2) \,.   
\end{equation}
We will assume that the integral converges. This imposes some weak constraints on the cutoff profile. From \eqref{F}, we see that we require  $C(q^2)$ to vanish slower than $q^{d+2}$ as $q\to0$, and $C\to1$ faster than $1/q^{d+2}$ as $q\to\infty$. This is true for example for the popular form of additive (\ie mass-type) cutoff \cite{Wetterich:1992} (which was the one used in the analogous $f(R)$ analyses in refs. \cite{Mitchell:2021qjr,Morris:2022btf}):
\begin{equation}
\label{wet}
    r(q^2)=\frac{q^2}{\text{exp}(aq^{2b})-1}\,, \quad a>0,\,b\geq 1 \,,
\end{equation}
provided also we set $b<\half(d+2)$, the relation to $C(q^2)$ being $q^2C^{-1}(q^2) = q^2 +r(q^2)$.

Given that $a_2(\ph)$ vanishes asymptotically, it is tempting to neglect the $a_2$ term in \eqref{eig}. We will shortly justify this. By neglecting the $a_2 $ term, the ODE becomes linear first order giving a unique solution up to normalization. Thus we deduce that the eigenoperators asymptotically scale as a power of the field:
\be \label{asympv} v(\ph) \propto |\ph|^{\frac{d-\lambda}{d_\ph}} +\cdots\,, \ee
where the ellipses stands for subleading corrections. 

The neglect of the $a_2$ is justified as follows.
The missing solution is one that grows exponentially (again, so that $a_2(\ph) v''(\ph)$ cannot be neglected). Since the ODE is linear, these are allowed solutions to \eqref{eig}, but they are ruled out because, on treating such perturbations at the non-perturbative level, it can be shown that they do not evolve multiplicatively in the RG no matter how close one starts to the FP \cite{Morris:1996nx,Morris:1998,Morris:1996xq,Bridle:2016nsu,Mitchell:2021qjr,Morris:2022btf} \ie the RG time dependence never takes the form in eqn. \eqref{eigenop}. (Such perturbations do not then have a well-defined scaling dimension, and in fact it can be shown that as soon as $\Lambda$ is lowered, they can be expanded as a convergent sum over the power-law solutions \eqref{asympv}. For more details, see refs. \cite{Morris:1996nx,Morris:1998,Morris:1996xq,Bridle:2016nsu,Mitchell:2021qjr,Morris:2022btf}.)

Now, the asymptotic solution \eqref{asympv} imposes two boundary conditions (one for each limit $\ph\to\pm\infty$) on the second order ODE \eqref{eig}, but since the ODE is linear this overconstrains the equation\footnote{We can see this for example by imposing a normalization condition on $v$.} which thus leads to quantisation of the RG eigenvalue $\lambda$. We index the solutions as $v_n(\ph)$, ordering them so that $\lambda_n$ decreases as $n$ increases. We can now perform an SL transformation and deduce the asymptotic dependence of the eigenvalues $\lambda_n$ on $n$, as $n\to\infty$.

\subsection{SL analysis}

We can rewrite the eigenvalue equation \eqref{eig} in a SL form by multiplying it with the SL weight function
\begin{equation} \label{weight}
    w(\ph)=\frac{1}{a_2(\ph)}\exp \left\{-\int_0^\ph d\ph' \frac{a_1(\ph')}{a_2(\ph')} d\ph' \right\} \,,
\end{equation}
which is always positive due to the positivity of $a_2$. Then the eigenvalue equation becomes
\begin{equation}
    -\left(a_2(\ph)w(\ph)v'(\ph) \right)'=(d-\lambda) w(\ph)v(\ph)\,. \label{SL}
\end{equation}
The SL operator on the left, $L=-\frac{d}{d\ph}\left( a_2 w \frac{d}{d\ph}\,\cdot \right)$, is self adjoint when acting on the space spanned by the eigenoperators, \ie it satisfies
\be \int^\infty_{-\infty}\!\!\!\!\!d\ph\, u_1(\ph)\, L u_2(\ph) = \int^\infty_{-\infty}\!\!\!\!\!d\ph\, u_2(\ph)\, L u_1(\ph)\,,\ee  
when the $u_i$ are linear combinations of the eigenoperators. This is so because the boundary terms at infinity, generated by integration by parts, vanish in this case. This follows because, from \eqref{asympv}, the $u_i$ diverge at worst as a power of $\ph$, whilst $w(\ph) \rightarrow 0$ exponentially fast as $\ph\rightarrow \pm \infty$. 

Thus from SL analysis \cite{Ince1956}, we know that the eigenvalues $\lambda_n$ are real, discrete, with a most positive (relevant) eigenvalue and an infinite tower of ever more negative (more irrelevant) eigenvalues, $\lambda_n\to-\infty$ as $n\to\infty$ \cite{Morris:1996xq}. Let us define a `coordinate' $x$:
\begin{equation}
    x = \int_0^\ph \frac{1}{\sqrt{a_2(\ph')}}\, d \ph' \label{x}
\end{equation}
(always taking the positive root in fractional powers).
Defining the wave-function as
\begin{equation}
\label{psi}
    \psi(x) = a_2^{1/4}(\ph)w^{1/2}(\ph)v(\ph)\,,
\end{equation}
enables us to recast \eqref{SL} as:
\begin{equation}
    -\frac{d^2\psi(x)}{dx^2}+U(x)\psi(x)=(d-\lambda)\psi(x) \,. \label{shrod}
\end{equation}
This is a one-dimensional time-independent Schrödinger equation for a particle of mass $m=1/2$, with energy $E=d-\lambda$ \ie just the eigenoperator scaling dimension, and with potential \cite{Benedetti:2013jk,Mitchell:2021qjr,Morris:2022btf}:
\begin{equation}
    U(x)=\frac{a_1^2}{4a_2}-\frac{a_1'}{2}+a_2'\left(\frac{a_1}{2a_2}+\frac{3a_2'}{16a_2}\right)-\frac{a''_2}{4}\,, \label{Ux}
\end{equation}
where the terms on the right hand side are functions of $\ph$. 

From the limiting behaviour of $a_2(\ph)$, \eqref{limA2}, we see that asymptotically the coordinate $x$ scales as
\begin{equation}
\label{largex}
    x = \int_0^\ph \left( \frac{|\ph'|^{m-2}}{\sqrt{F}} + O(1) \right) d\ph' = \pm \frac{|\ph|^{m-1}}{(m-1)\sqrt{F}} +O(|\ph|) \quad \text{as} \quad \ph \rightarrow \pm\infty \,,
\end{equation}
so in particular when $\ph \rightarrow \pm \infty$ we have $x\rightarrow \pm \infty$.
On the right hand side of \eqref{Ux}, the first term dominates at leading order (LO) and next-to-leading order (NLO). Since asymptotically,
\be \frac{a_1^2(\ph)}{4a_2(\ph)}=\frac{d_\ph^2}{4F}|\ph|^{2m-2} +O(|\ph|^m)\,, \ee
we thus find that
\begin{equation} \label{Uasymp}
    U(x) = \frac{1}{4}(d-d_\ph)^2 x^2 +O(|x|^{1 + \frac{1}{m-1}}) \quad \text{as} \quad x \rightarrow \pm \infty \,.
\end{equation}
To LO, this is the potential of a simple harmonic oscillator of the form $\frac{1}{2}m\omega^2x^2$, where 
\be \label{omega}\omega = d-d_\ph = \frac12(d+2-\eta)\,. \ee

\subsection{WKB analysis}
We can now use WKB analysis to compute the asymptotic form of the energy levels, \aka operator scaling dimensions, $E_n$, at large $n$. This follows from  solving the equality
\begin{equation} \label{WKB1}
    \int_{-x_n}^{x_n}\!\!\!\! dx\, \sqrt{E_n - U(x)} = \left( n + \frac{1}{2} \right) \pi \,,
\end{equation}
for the total phase of the wave oscillations described by $\psi(x)$,
in the limit of large $E_n$ \cite{morse1953methods}. Here $x_n$ are the classical turning points, \ie such that $E_n = U(\pm \, x_n)$. Now, the above integral is dominated by the regions close to the turning points, where 
we can substitute the asymptotic form  \eqref{Uasymp}. Including the subleading correction proportional to some constant $\gamma$ (that depends on the cutoff profile) the integral is
\begin{equation} \label{WKB2}
    \frac{\omega}{2} \int_{-x_n}^{x_n}\!\!\!\! dx\, \sqrt{x_n^2 + \gamma x_n^{1 + \frac{1}{m-1}} - x^2 - \gamma |x|^{1 + \frac{1}{m-1}}} = 
    \frac{\omega}{2}x^2_n \int_{-1}^{1}\!\!\!\! dy\,\sqrt{1-y^2+\gamma x_n^{\frac{1}{m-1}-1}(1-|y|^{1+ \frac{1}{m-1}})} \,.
\end{equation}
Since the $x_n$ are also large we can now evaluate the right hand side and thus from \eqref{WKB1} we get the asymptotic relation between $x_n$ and $n$:
\begin{align}
    \frac{\omega \pi}{4} x_n^2 + O\left( x_n^{1+\frac{1}{m-1}} \right) = n \pi\,. \label{WKB8}
\end{align}
Hence, using \eqref{Uasymp}, \eqref{omega} and \eqref{WKB8}, the scaling dimension of the eigenoperators takes the form
\be 
\label{largen}
d_n=E_n=d-\lambda_n=U(x_n) = n\omega + O\left( n^\frac{m}{2(m-1)} \right) = n(d-d_\ph) + O\left( n^\frac{m}{2(m-1)} \right) \quad \text{as} \quad n \rightarrow \infty \,.
\ee
The subleading correction to the critical exponents contain information about the cutoff via the constant $\gamma$ introduced in \eqref{WKB2}. However, at leading order the result is independent of the cutoff, and is hence universal.

\section{O(N) scalar field theory}
\label{sec:on}

Now let us apply the same treatment to $N$ scalar fields $\varphi^a$ ($a = 1,\ldots,N$) with an $O(N)$ invariant potential $V_\Lambda(\varphi^2) = V_\Lambda(\rho)$, in the LPA. We use the shorthand $\rho = \varphi^a \varphi^a = \varphi^2$.  The flow equation \eqref{flowLPA} becomes \cite{Morris:1997xj,DAttanasio:1997he}:
\begin{equation} \label{flowN3}
    \left( \partial_t - d + 2 d_\varphi \rho \frac{\partial}{\partial \rho} \right) V_\Lambda(\rho) = - \frac{1}{2} \int \frac{d^d q}{(2\pi)^d} \frac{\dot{\Delta}}{\Delta}  \left( M^{-1} \right)^{aa} \,, 
\end{equation}
where the matrix $M$ is given by:
\begin{equation} \label{M2}
    M^{ab} = \delta^{ab}+\Delta \frac{\partial^2 V_\Lambda(\rho)}{\partial \varphi^a \partial \varphi^b} = \delta^{ab} + 2 \Delta \left[ \delta^{ab} V_\Lambda'(\rho) + 2 \varphi^a \varphi^b V_\Lambda''(\rho) \right] \,.
\end{equation}
Inverting and tracing, yields:
\begin{equation} \label{trace}
    \left( M^{-1} \right)^{aa} = \frac{N-1}{1 + 2 \Delta V_\Lambda'(\rho)} + \frac{1}{1 + 2 \Delta V_\Lambda'(\rho) + 4 \Delta \rho V_\Lambda''(\rho)} \,.
\end{equation}
In the limit of large $\rho$, the right hand side of the flow equation \eqref{flowN3} can be neglected at leading order. This implies that a FP solution $V_\Lambda(\rho) = V(\rho)$ takes the following asymptotic form:
\begin{equation} \label{FPN}
    V(\rho) = A \rho^{\frac{m}{2}} + O\left(\rho^{1-\frac{m}{2}}\right) \quad \text{as} \quad \rho \rightarrow \infty \,,
\end{equation}
where as before the subleading term has been calculated by iterating the leading contribution to next order.

The RG eigenvalue equation follows by linearising \eqref{flowN3} around the fixed point solution,
\begin{equation} \label{linearN}
    V_\Lambda(\rho) = V(\rho) + \varepsilon\, v(\rho)\, \e^{\lambda t} \,, 
\end{equation}
giving an equation for $v(\rho)$ with the same structure as \eqref{eig}, \ie
\begin{equation} \label{eigN}
    -a_2(\rho) v'' + a_1(\rho) v' + a_0(\rho) v = (d-\lambda) v \,,
\end{equation}
the same value for $a_0(\rho) = 0$, but different expressions for $a_1(\rho)$,
\begin{equation} \label{a1N}
    a_1(\rho) = 2 d_\varphi \rho - \int \frac{d^d q}{(2\pi)^d} \dot{\Delta} \left[ \frac{1}{\left( 1 + 2 \Delta V' + 4 \Delta \rho V'' \right)^2} + \frac{N-1}{\left( 1 + 2 \Delta V' \right)^2} \right]\,,
\end{equation}
and $a_2(\rho)$, which however is again always positive:
\begin{equation} \label{a2N}
    a_2(\rho) = \int \frac{d^d q}{(2\pi)^d}  \frac{2\dot{\Delta}\rho}{\left( 1 + 2 \Delta V' + 4 \Delta \rho V'' \right)^2}\,.
\end{equation}
Using the asymptotic fixed point solution \eqref{FPN} (and assuming $A\ne0$) we get that asymptotically $a_2$ scales as follows: 
\begin{equation}
\label{a2Nasymp}
    a_2(\rho) = 4F\rho^{3-m}  +O\left(\rho^{4-\frac{3m}{2}}\right) \quad \text{as} \quad \rho \rightarrow \infty \,,
\end{equation}
where $F$ was already defined in \eqref{F}. By similar arguments to before, we see that $m>3$ in practice, so this implies $a_2(\rho)\to0$.
We also find that $a_1$ scales as follows:
\begin{equation}
\label{a1Nasymp}
    a_1(\rho) = 2\, d_\varphi \rho + O\left(\rho^{2-m}\right) \quad \text{as} \quad \rho \rightarrow \infty \,.
\end{equation}
If we substitute $\rho=\ph^2$ into the above asymptotic expansions, they differ from the large $\ph$ behaviour \eqref{a1} of $a_1(\ph)$ and \eqref{limA2} of $a_2(\ph)$. However they reproduce the previous results once we transform the ODE \eqref{eigN} by changing variables $\rho=\ph^2$. Thus by the same arguments as before, \cf \eqref{asympv}, we also know that for $\rho\to\infty$, we must have
\be \label{asympvN} v(\rho) \propto \rho^{\frac{d-\lambda}{2d_\ph}} +\cdots\,. \ee
However, this now imposes only one boundary condition on the linear ODE \eqref{eigN} since $\rho$ is restricted to be non-negative. On the other hand we see from \eqref{a2N} that $a_2(0)=0$, so the ODE has a so-called fixed singularity at $\rho=0$. In order to ensure that $v(\rho)$ remains non-singular at this point, an additional boundary condition is then required:
\be\label{bc0}  a_1(0)v'(0)=(d-\lambda) v(0) \,.\ee 
Now we again have two boundary conditions, overconstraining the equation, and leading to quantisation of the RG eigenvalue $\lambda$.

\subsection{SL analysis}

The last step is to perform the SL analysis, which also differs because of the $\rho=0$ boundary. For small $\rho$ we have 
\be \label{asm} a_2(\rho) =2G\rho+O(\rho^2)\qquad\text{and}\qquad a_1(\rho) =-GN+O(\rho)\,, \ee
where we have set
\be \label{G}
G= \int \frac{d^d q}{(2\pi)^d} \frac{\dot{\Delta}}{\left[ 1 + 2 \Delta V'(0) \right]^2}  \,.
\ee
Note that $G$ is of course positive. (By Taylor expanding \eqref{flowN3} one sees that its convergence is guaranteed for any such solution to the flow equation.) 
The SL weight function now takes the form
\begin{equation} \label{weight_N}
    w(\rho) = \frac{1}{a_2(\rho)} \exp \left\{ - \int^\rho_{\rho_0}\!\!\!\! d\rho'\, \frac{a_1(\rho')}{a_2(\rho')} \right\} \,,
\end{equation}
where by \eqref{asm} a non-zero lower limit, $\rho_0>0$, is required to avoid the integral diverging (when $N\ne0$).

Using $w(\rho)$ we can now cast \eqref{eigN} in SL form \eqref{SL}. However, for the SL operator to be self-adjoint, we need the boundary contributions that appear on integration by parts, to vanish. This is still true for large field since as $\rho\to\infty$, the eigenoperators diverge at worst as a power, whilst  from \eqref{a2Nasymp} we have $a_2(\rho)\to0$, and thus $w(\rho)\to0$ exponentially fast. At the $\rho=0$ boundary we require:\footnote{Using  \eqref{bc0} and \eqref{asm}, this can be reduced to $\lim_{\rho\to0}a_2(\rho)w(\rho)(\lambda_i-\lambda_j)\,v_i(\rho)v_j(\rho) = 0$ (when $N\ne0$).}
\begin{equation}
\lim_{\rho\to0}  a_2(\rho)w(\rho)\left(v_i(\rho)v_j'(\rho)-v_j(\rho)v'_i(\rho)\right)=0 \,, \label{w(0)}
\end{equation}
for any two eigenfunctions $v_i(\rho)$ and $v_j(\rho)$. This is true for all $N>0$ since by \eqref{asm} and \eqref{weight_N} we see that for small $\rho$,
\be  
\label{a2wsmall}
a_2(\rho)w(\rho) \propto \rho^{N/2} \left[1+O(\rho)\right]\,.
\ee
We have thus determined that the SL operator is self-adjoint for all $N>0$. 

Actually, $N=0$ is also interesting since it corresponds to the universality class of fluctuating long polymers \cite{ZinnJustin:2002ru}. In this case, the above analysis shows that $a_2(0)w(0)>0$, which would appear to imply that \eqref{w(0)} is no longer satisfied. However from \eqref{asm} we see that $a_1(0)=0$ now and thus, from \eqref{bc0}, either $\lambda_i=d$ or $v_i(0)=0$ \cite{Morris:1997xj}. The first possibility corresponds to the uninteresting solution $v(\rho)\equiv1$, \ie the unit operator, which we discard. All the other eigenoperators must thus satisfy $v_i(0)=0$, and so \eqref{w(0)} is satisfied in this reduced space. Therefore, with this one proviso, the SL operator is actually self-adjoint for all $N\ge0$. 

For general $N<0$, the SL operator fails to be self-adjoint, and thus SL analysis is no longer applicable. However for $N=-2k$, $k$ a non-negative integer,  something special happens. The first $k+1$ eigenoperators with the lowest scaling dimension turn out to have exactly soluble scaling dimensions, in fact coinciding with the Gaussian ones \cite{PhysRevLett.30.544,PhysRevLett.30.679,RevModPhys.46.597}. (The case $N=0$ above is the first example, the lowest dimension operator being the unit operator with scaling dimension zero.) Again, the SL operator is self-adjoint in the remainder of the space. For example for $N=-2$, one knows from ref. \cite{Morris:1997xj} that the remaining eigenoperators satisfy $v_i(0)=v'_i(0)=0$, and thus $v_i(\rho)\propto\rho^2$ for small $\rho$, whilst for $N=-4$ boundary conditions force the remaining eigenoperators to satisfy $v_i(\rho)\propto \rho^3$ for small $\rho$. From that analysis it is clear that in general at $N=-2k$, we have that the remaining operators satisfy 
\be \label{vmcase} v_i(\rho)\propto\rho^{k+1}\qquad\text{as}\qquad\rho\to0\,.\ee 
Combining these observations with \eqref{w(0)} and \eqref{a2wsmall}, we see that the SL operator is indeed self-adjoint in the reduced space defined by excluding the first $k+1$ operators.

The SL equation can now be recast in the same way as before, using \eqref{x} for $x$  and \eqref{psi} for $\psi(x)$ (except for the obvious replacement of $\ph$ by $\rho$). The resulting Schr\"odinger equation is then precisely as before, \viz \eqref{shrod}, and the potential $U(x)$ also takes precisely the same form in terms of the $a_i$, \viz \eqref{Ux}. 
However the $\rho=0$ boundary turns into an $x=0$ boundary since, by \eqref{asm} and \eqref{x}, we have
\begin{equation} \label{xsmall}
    x = \sqrt{2\rho/G} + O\left( \rho^\frac{3}{2} \right) \quad \text{as} \quad \rho \rightarrow 0 \,.
\end{equation}
Thus, using $a_2$ from \eqref{asm} and $a_2w$ from \eqref{a2wsmall}, we see that
\be\label{psism} \psi(x)\propto x^{\frac{N-1}{2}}v(x)\ee
for small $x$. Hence for all $N>1$, $\psi(x)$ vanishes as $x\to0$.  On taking into account the behaviour \eqref{vmcase} we see that in the reduced space, $\psi(x)$ also vanishes for the special cases $N=-2k$. In this limit the leading contributions to the potential come from the first, third and fourth terms in \eqref{Ux}, and thus we find:
\begin{equation} \label{Usm}
    U(x) = \frac{(N-1)(N-3)}{4\,x^2} +O(1)
 \quad \text{as} \quad x \rightarrow 0 \,.
\end{equation}
The cases $N=1,3$ are exceptional since this leading behaviour then vanishes, whilst the range $1<N<3$ will need a separate treatment because the potential is then unbounded from below.  


At the other end of $x$'s range, we find that 
\begin{equation} \label{xN}
    x = \int_0^\rho\!\!\! d\rho' \left( \frac{\left( \rho' \right)^{\frac{1}{2}(m-3)}}{2 \sqrt{F}} + O\left( \rho'^{-\frac{1}{2}} \right) \right) = \frac{\rho^{\frac{1}{2}(m-1)}}{(m-1) \sqrt{F}} + O\left( \rho^{\frac{1}{2}}\right) \quad \text{as} \quad \rho \rightarrow \infty \,.
\end{equation}
Identifying $\rho=\ph^2$, this is the same formula \eqref{largex} as before. The potential $U(x)$ is again dominated by the first term in \eqref{Ux}, both at LO and NLO. Substituting the asymptotic expressions \eqref{a1Nasymp} and \eqref{a2Nasymp} for $a_1$ and $a_2$, we find exactly the same formula \eqref{Uasymp} for the large $x$ behaviour of $U(x)$. In particular the leading term is again that of a simple harmonic oscillator with angular frequency $\omega = d-d_\ph$.

\subsection{WKB analysis}
 
For the cases $N>3$, $0<N<1$ and $N=-2m$, we can now proceed with the WKB analysis in the usual way. In this case we have for the total phase of the wave function:
\begin{equation} \label{WKB_N}
    \int_{x^-_n}^{x^+_n} \!\!\!\! dx\, \sqrt{E_n - U(x)} = \left( n + \frac{1}{2} \right) \pi \,,
\end{equation}
where $x^-_n$ and $x^+_n$ are the classical turning points, \ie $E_n=d-\lambda_n=U(x^-_n)=U(x^+_n)$. In contrast to the previous case, the potential is not symmetric and there is no simple relation between $x^-_n$ and $x^+_n$. 

In the large $n$ limit, the contribution from the right hand boundary gives half of what we obtained before.  To see this in detail, let $x_0^+$ be some fixed finite value but sufficiently large to trust the asymptotic form \eqref{Uasymp} of the potential, then the contribution from the right hand boundary is
\be 
\int_{x^+_0}^{x^+_n} \!\!\!\! dx\, \sqrt{E_n - U(x)} = \frac{\omega}{2}(x^+_n)^2 \int_{x^+_0/x^+_n}^{1}\!\!\!\! dy\,\sqrt{1-y^2+\gamma (x^+_n)^{\frac{1}{m-1}-1}(1-|y|^{1+ \frac{1}{m-1}})}\,.
\ee
Taking into account the multiplying factor of $(x^+_n)^2$ we see that the lower limit $x^+_0/x^+_n$ of the integral can be set to zero, since the correction is of order $O(x^+_n)$ which is smaller than that given by the $\gamma$ correction. Thus we get half the integral in \eqref{WKB2} (with $x_n$ replaced by $x^+_n$) giving half the left hand side of \eqref{WKB8}:
\be 
\label{WKBR}
\int_{x^+_0}^{x^+_n} \!\!\!\! dx\, \sqrt{E_n - U(x)} =  \frac{\omega \pi}{8} (x^+_n)^2 + O\left( (x^+_n)^{1+\frac{1}{m-1}}\right)\,.
\ee
Using the asymptotic form of the potential, we see that the leading term can be written as $\pi E_n/(2\omega)$.
In the large $n$ limit, the left hand boundary makes a contribution that can be neglected in comparison. To see this let $x^-_0$ be some fixed finite value but sufficiently small to use \eqref{Usm}. Then the contribution from the left hand boundary is
\be 
\int_{x^-_n}^{x^-_0} \!\!\!\! dx\, \sqrt{E_n - U(x)} = \frac12\sqrt{(N-1)(N-3)}\int_1^{x^-_0/x^-_n}\!\!\!\! dy\,\left(\frac{\sqrt{y^2-1}}{y}+O(x^-_n)\right)\,.
\ee
Since $x^-_n$ is vanishing for large $E_n$, we see that this integral is $O(1/x^-_n)$ or, using again the relation \eqref{Usm}, $O(E^{1/2}_n)$. 
That only leaves the portion of the integral that goes from $x^-_0$ to $x^+_0$, but since these boundaries are fixed and finite, we see that this part also grows as $\sqrt{E_n}$ and thus it too can be neglected in comparison to \eqref{WKBR}.

Therefore asymptotically the integral in \eqref{WKB_N} is given by \eqref{WKBR}. Inverting the relation to find $(x^+_n)^2$ asymptotically in terms of $n$, we thus find 
\be
\label{largenN}
d_n=E_n=d-\lambda_n=U(x^+_n) = 2n\omega + O\left( n^\frac{m}{2(m-1)} \right) = 2n(d-d_\ph) + O\left( n^\frac{m}{2(m-1)} \right) \quad \text{as} \quad n \rightarrow \infty \,,
\ee
\ie precisely double the value we found for a single component field in \eqref{largen} and independent of $N$.

We see that technically this arises because the WKB integral is precisely half as large in the $O(N)$ case, the leading contribution coming from the $x^+_n$ boundary only. Recall that at $N=1,3$, the leading behaviour \eqref{Usm} of $U(x)$ is no longer applicable. Since the potential is now finite as $x\to0$, it is clear from the above analysis that the left hand boundary continues to contribute at most $O(E^{1/2}_n)\sim\sqrt{n}$ and so can be neglected. Thus we see that \eqref{largenN} applies also to these exceptional cases. Thus also for $N=1$ we find twice the previous scaling dimension as a function of large index $n$. This is in agreement with that single field result however, because these eigenoperators are a function of $\ph^2$ only. Hence for a single component field, the current $n$ indexes only the even eigenoperators (those symmetric under $\ph\leftrightarrow-\ph$).

Finally, let us show that our result \eqref{largenN} is also applicable to the range $1<N<3$.  Although in this case, from \eqref{Usm}, the potential $U(x)\to-\infty$ as $x\to0$, we know from \eqref{psism} that the solutions we need, have $\psi(x)$ vanishing there. These solutions are consistent with the Schr\"odinger equation \eqref{shrod} because for small $x$ we have, by \eqref{psism}, a diverging second derivative:
\be -\frac{d^2\psi(x)}{dx^2} \propto -\frac{(N-1)(N-3)}{4\,x^2}\psi(x)\,,\ee
which is precisely the right behaviour to cancel the divergence in the Schr\"odinger equation coming from the $U(x)\psi(x)$ term. Meanwhile the $v(x)$ term in \eqref{psism} is well behaved in terms of oscillations at small $x$, behaving similarly to the above cases. Therefore we are only neglecting a subleading contribution to the total phase, if we work instead with a modified WKB integral where we replace the lower limit in \eqref{WKB_N} with some finite value $x^-_0$. By the above analysis we then recover \eqref{largenN} again. In this way we have shown that the result \eqref{largenN} is actually applicable for all $N\ge0$ and to the special cases $N=-2k$ (where $k$ is a non-negative integer).

%
%
%
%

\section{Summary and discussion}
\label{sec:summary}

We have used SL theory and WKB  methods to derive the scaling dimension $d_n$ of highly irrelevant operators $\cO_n$ around a non-trivial fixed point for scalar field theory, in the Local Potential Approximation (LPA).
The scaling dimensions $d_n$ are ordered so that they increase with increasing index $n$. The $d_n$ are derived following the methods developed in \cite{Mitchell:2021qjr}. They are given to leading order in $n$, together with the power-law dependence on $n$ of the next-to-leading order. The results apply to all the non-trivial (multi)critical fixed points in $2<d<4$, for single component scalar field theory and for $O(N)$ invariant scalar field theory, and also to the unitary minimal models in $d=2$ dimensions.
The $d_n$ are universal, independent of the choice of fixed point (except through the anomalous dimension $\eta$) and independent of the cutoff choice which we have left general throughout, apart from the weak technical constraints discussed below eqn. \eqref{F}. In particular these constraints allow for the popular smooth cutoff choice \eqref{wet}. The crucial property leading to universality is that the results depend only on asymptotic solutions at large field, which can be derived analytically, and are also universal in the same sense. Although non-universal cutoff-dependent terms, in particular \eqref{F} and \eqref{G}, enter into the calculation at intermediate stages, they drop out in the final stages. For a single component real scalar field, $d_n$ is given in \eqref{largen}. 
For  $O(N)$ scalar field theory, the $d_n$ are just twice this, \cf \eqref{largenN}, independent of $N$. This is in agreement with the single field result because here $n$ indexes the eigenoperators that are a function of $\ph^2$ only.

The first steps in deriving these results is to recast the eigenoperator equation in SL form, and then establish that the SL operator is self-adjoint in the space spanned by the eigenoperators. For a single component scalar field this follows after demonstrating that the SL weight decays exponentially for large field, since the eigenoperators grow at most as a power of the field. For the $O(N)$ case the analysis is more subtle because the relevant space is now the positive real line (parametrised by $\rho=\ph^2\ge0$) and thus the SL operator is self-adjoint only if the boundary terms at $\rho=0$ also vanish. By analytically determining the small $\rho$ dependence of the relevant quantities we see that the SL operator is self-adjoint when $N>0$. For $N\le0$, the SL operator is not self-adjoint and the analysis does not apply. Presumably in these cases one would find that the scaling dimensions $d_n$ are no longer real. However for a sequence of special cases $N=-2k$, $k$ a non-negative integer, the SL operator is self-adjoint on a reduced space spanned by all eigenoperators apart from the first $k+1$. The analysis can then proceed on this reduced space. As we already noted, while most of these special cases are presumably only of theoretical interest, the $N=0$ case describes the statistical physics of long polymers.

The next step is to cast the SL equation in the form of a one-dimensional time-independent Schr\"odinger equation with energy levels $E_n=d_n$ and potential $U(x)$. For the single component field this potential is symmetric, and in order to determine the energy levels $E_n$ asymptotically at large $n$, using the WKB approximation, we need only the behaviour of $U(x)$ at large $x$. The latter follows  from our asymptotic analysis. For $O(N)$ scalar field theory, the space is the positive real line $x\ge0$, and thus for WKB analysis we need also the behaviour of the potential $U(x)$ at small $x$. Here we find that the range $1\le  N\le 3$ requires a separate treatment because the leading term in $U(x)$ turns negative leading to a potential unbounded from below. Nevertheless we are able to treat this case and the end result for $d_n$, \eqref{largenN}, is the same, thus applying universally to all $N\ge0$ and the $N=-2k$ special cases.

Although these results are universal, they are still derived within the LPA, which is an uncontrolled model approximation. One might reasonably hope however that the fact that these results are universal in the sense of being independent of the detailed choice of cutoff, is an indication that they are nevertheless close to the truth. On the other hand the LPA \cite{Ball:1994ji} of the Polchinski flow equation \cite{Polchinski:1983gv} is in fact completely cutoff independent, although this property arises rather trivially. It is actually equivalent under a Legendre transformation \cite{Morris:2005ck} to the flow equation \eqref{flowLPA} for the Legendre effective action in LPA, as we study here, but only for a special (but actually popular) choice of additive cutoff known as the optimised cutoff \cite{opt1}. However the optimised cutoff does not satisfy our technical constraints given below \eqref{F} so our analysis is invalid for this case. Nor in fact does a sharp cutoff \cite{NCS,Hasenfratz:1985dm,Morris:1993,Morris:1994ki} or power-law cutoff \cite{Morris:1994ie} satisfy the technical constraints. What this means is that these particular cutoffs fail to regularise completely the region of large fields, in the sense that $a_2$, defined by \eqref{a2} or \eqref{a2N}, no longer has an asymptotic expansion given simply by integrating over the asymptotic expansion of its integrand. For these three particular cutoffs, regions of momenta far from $\Lambda$ alter the
asymptotic expansion of $a_2$ so that it is no longer of the form \eqref{limA2}, or \eqref{a2Nasymp}, and for this reason these cutoffs are less satisfactory.  

Nevertheless, following our methods, it would be straightforward to derive the asymptotic scaling dimensions $d_n$ in LPA for any or all of these three special choices of cutoff, by using the particular form of the LPA flow equation in these cases (which are known in closed form, since the momentum integrals can be calculated analytically in these cases). The results will differ from the $d_n$ derived here and amongst themselves, but their investigation would improve insight into the accuracy of the LPA in this regime. Furthermore it would seem possible to generalise any of these special choices of cutoff to their own class of cutoffs with similar properties, and thus understand the extent to which the results could still be cutoff independent, up to some appropriate constraints, in these cases, and gain a more detailed understanding of why the $d_n$ differ.

Unfortunately our $d_n$ do not seem to match in a useful way to existing results in the literature. The LPA restricts us to eigenoperators that contain no spacetime derivatives, and thus our index $n$ counts only over these. In reality all eigenoperators (apart from the unit operator) contain spacetime derivatives, so in particular it is not clear how our index $n$ would map into the exact sequence.

However in some special limits the LPA is effectively exact. This is true for the Gaussian fixed point for example, where $d_n = nd_\ph$ (with $\eta=0$). Our scaling dimensions $d_n$ differ from this, but the Gaussian fixed point is specifically excluded from our analysis since our results apply only to non-trivial fixed points, such that the asymptotic expansion of the fixed point potential takes the form \eqref{asympV} or \eqref{FPN} with $A\ne0$. 

The LPA also becomes effectively exact in the large $N$ limit \cite{DAttanasio:1997he}, and there the scaling dimensions are $d_n=2n$ (with $\eta=0$) which again differs from our result (as well as differing from the Gaussian fixed point result). Furthermore they continue to disagree even if we now take a second limit such that both $n$ and $N$ are sent to infinity. However in this case we have an example where the order of the limits matters. The $N\to\infty$ result is derived for $d_n$ whilst first holding $n$ fixed, while our result applies first for fixed $N$ while $n\to\infty$. 

The difference can be seen at the technical level. The first term on the right hand side of the flow equation \eqref{flowN3} is proportional to $N$. In our analysis however it is the denominators that dominate. On the other hand in the large $N$ analysis, only the first term survives, resulting in a first order ODE with no SL properties (or Schr\"odinger equation representation). The universal results fall out on the one hand in our analysis from the asymptotic behaviour at large field, but on the other hand in large $N$ they fall out from a Taylor expansion around the minimum of the fixed point potential \cite{DAttanasio:1997he}. There seems unfortunately to be no way to bridge the gap between these two limiting regimes.

An even clearer example where the exchange of limits do not commute, is provided by the special cases $N=-2k$. As we recalled in sec. \ref{sec:on}, in these cases the first $k+1$ eigenoperators degenerate, gaining Gaussian scaling dimensions. But our $d_n$ apply to the highly irrelevant eigenoperators that are found in the reduced space, which excludes these first $k+1$ operators, and hence have non-trivial scaling dimensions. However if instead we fix on the $n^\text{th}$ eigenoperator and let $N\to-\infty$ by sending $k\to\infty$, we see that this $n^\text{th}$ eigenoperator will fall into the excluded space and thus end up with Gaussian scaling dimensions. The disagreement between the two results will then remain even if we choose next to send $n\to\infty$.

 \section{Acknowledgements}
VMM and DS acknowledge support via STFC PhD studentships. TRM acknowledges support from STFC through Consolidated Grant ST/T000775/1.


\bibliographystyle{hunsrt}
\bibliography{references.bib}


\end{document}